\documentclass[conference]{IEEEtran}
\IEEEoverridecommandlockouts
\usepackage{cite}
\usepackage{algorithmic}
\usepackage{graphicx}
\usepackage{textcomp}
\usepackage{xcolor}
\usepackage{array}
\usepackage{bbm}
\usepackage[hidelinks]{hyperref}
\usepackage{xcolor}
\hypersetup{
	colorlinks  = true,
	linkcolor	={red!100!black},
	citecolor	={blue!100!black},
	urlcolor	={blue!100!black}
}
\usepackage{mathtools}
\usepackage{subfigure,epsfig}
\usepackage{amsmath, amsthm, amssymb}
\usepackage[outdir=./]{epstopdf}
\usepackage{dsfont}
\usepackage{mathrsfs}
\newtheorem{Theorem}{Theorem}
\newtheorem{Lemma}{Lemma}

\def\figref#1{Fig.\,\ref{#1}}%

\def\BibTeX{{\rm B\kern-.05em{\sc i\kern-.025em b}\kern-.08em
    T\kern-.1667em\lower.7ex\hbox{E}\kern-.125emX}}
\begin{document}

\title{On the Effect of Temporal Correlation on Joint Success Probability and Distribution of Number of Interferers in Mobile UAV Networks \\
}

\author{Mohammad Salehi and Ekram Hossain\thanks{The authors are with the 
Department of Electrical and Computer Engineering, 
University of Manitoba, Canada (Email: salehim@myumanitoba.ca, Ekram.Hossain@umanitoba.ca). This work was supported by a Discovery Grant from the Natural Sciences and Engineering Research Council of Canada (NSERC).}}	



\maketitle

\begin{abstract}
 Unmanned aerial vehicle (UAV)-aided terrestrial cellular networks is being considered as one of the disruptive technologies that can address the beyond 5G (B5G)/6G requirements. With mobile UAVs, since locations of the serving  UAVs and the interfering UAVs can vary over time, interference and other performance metrics become temporally correlated. 
In this letter, we analyze the effect of temporal correlation on the joint success probability and the distribution of the number of interferers for mobile UAV networks. The analytical results will be useful, for example,  to design error recovery protocols in these networks.   
\end{abstract}


\section{Introduction and Motivation}
To meet the B5G and 6G requirements, UAVs can be employed to complement the terrestrial base stations to provide wireless communications service to the mobile users. 
Depending on the system model, different point processes from stochastic geometry have been used in the existing literature to model a UAV network and study downlink communication performance in different scenarios (e.g. in \cite{Turgut}, \cite{Galkin2017ASG,Azari2017,Kim2018}, \cite{Sharma2019}). 
However, all of the existing work on static and mobile UAV scenarios only consider one time slot. Since locations of UAVs are spatially and temporally correlated, interference and other performance metrics that depend on the locations of the nodes are also spatially (at different locations in the network) and temporally (at different time slots) correlated \cite{ganti2009spatial}. Therefore, results that are derived for one time slot cannot be easily extended to study the network performance over time.


In this letter, we study the effect of correlation on the {\em joint success probability} (i.e. the joint probability of success at two time instants) and the {\em distribution of the number of interferers} in a mobile UAV network. The motivation for this study is as follows: (i) At a typical receiver, success probability is the probability that the signal-to-interference-plus-noise ratio (SINR) is greater than a target threshold. Due to correlation, unsuccessful transmission at time $0$ (reference time) implies that probability of successful retransmission at time $t$ tends to $0$ as $t\to 0$, i.e. probability of successful retransmission at time $t$ is not independent of the success probability at time 0. Therefore, the joint transmission probability can be exploited to design error recovery (e.g. ARQ [Automatic Repeat reQuest]) strategies. (ii) Given that $m$ co-channel UAVs cause interference at the typical user at time $0$, the number of interferers at time $t$ depends on $m$, the initial locations of all UAVs, mobility parameters (velocity and direction) of all UAVs, and UAVs' antenna patterns. With the distribution of the number of interferers at time $t$ given the number of interferers at time $0$ is $m$, we can study the success probability-latency tradeoff. For example, in some scenarios (depending on the network parameters), postponing a transmission for time $t$ can increase the success probability. {\em Note that we only consider the effect correlation at the typical user at two different time instants (i.e. temporal correlation)}.

The rest of the letter is organized as follows. Section II states the system model and assumptions. The analytical results on joint success probability and distribution of the number of interferers are presented in Section III.  Section IV presents the numerical and simulation results and Section V concludes the letter.

\section{System Model and Assumptions}

We assume all UAVs are positioned at the same height $h$ and projection of their initial locations onto $\mathbb{R}^2$ plane follows a homogeneous Poisson point process (PPP) $\Phi_{0}$ of intensity $\lambda$. Due to the stationarity of PPP, we can randomly select a typical UAV and consider its projected location as the origin of our coordination system $o\equiv(0,0,0)$. We assume each UAV is serving a user that is located beneath it at the ground level, i.e. UAVs are located such that the average received signal power at each user from its serving UAV is maximum. Thus, the typical user which is served by the typical UAV is located at the origin $o$. Since $h$ is fixed, for the rest of the paper, when we discuss about UAVs' locations we mean their projected locations onto the $\mathbb{R}^2$ plane. Therefore, when we say a UAV is located at $x\in\mathbb{R}^2$, it means the projected location of the UAV onto the $\mathbb{R}^2$ plane is at $x$ and the UAV is at distance $\sqrt{\|x\|^2+h^2}$ from the typical user, where $\|x\|$ denotes the 2-norm distance.

We assume that, after the first time slot, a fraction $p$ of UAVs (except the typical UAV) are mobile, i.e. $p$ denotes the probability that a UAV moves which is independent of its initial location and other UAVs' locations. A mobile UAV that is initially located at $x$, moves, for duration $t$, with velocity $v_x$ in a random angle $\theta_x$ with respect to the direction to the typical UAV, where $\theta_x$ is uniformly distributed in the interval $[0,2\pi]$ and $v_x$ is a random variable with probability density function (PDF) $f_V(v)$ and cumulative distribution function (CDF) $F_V(v)$. We also denote UAV's displacement due to mobility by $\vec{\text v}_xt$; therefore, 
The UAV's projected location at time $t$ can be written as $x+\vec{\text v}_xt$. We use $\Phi_t$ to denote the projected locations of UAVs at time $t$.

At time $t$, channel between a UAV located at $x$ and the typical user at the origin is  $h_x(t)\ell(\|x\|)$, where $h_x(t)$ denotes the channel power coefficient and follows a gamma distribution with shape parameter $k$ and scale parameter $\Omega$ (Nakagami fading) with the PDF is as follows:
\begin{IEEEeqnarray}{rCl}
	f_h(x) = \frac{1}{\Gamma(k)\Omega^k}x^{k-1}e^{-{x}/{\Omega}}. \nonumber 
\end{IEEEeqnarray}
We assume that $h_x(t)$ is independently and identically distributed across time $t$ and UAV location $x$. $\ell(\|x\|)=\|x\|^{-\alpha}$ also denotes the large-scale path loss, where $\alpha>2$ is the path-loss exponent. We also denote the variance of thermal noise by $\sigma^2$.

UAV antenna downtilt angle $\theta_{\rm t}$ is $\pi/2$. We also use sectorized gain pattern to approximate UAV antenna gain. Specifically, we assume
\begin{IEEEeqnarray}{rClrCl}
	G(\theta) &=& 
	\begin{cases}
		G_{\rm m}  \qquad \; \; \; \;  \; \; \: \:                    |\theta| \le {\theta_{\rm m}} \\  
		G_{\rm s}  \qquad                  {\theta_{\rm m}} < |\theta| \le {\theta_{\rm s}} \\   
		0          \qquad \; \; \; \; \; \; \:                 \text{Otherwise}
	\end{cases},
	\nonumber
	\label{eq:antenna_gain}
\end{IEEEeqnarray}
i.e. antenna gain is $G_{\rm m}$ for all the angles within the main lobe of beam width $2\theta_{\rm m}$ and is $G_{\rm s}$ for all the angles between $\theta_{\rm m}$ and $\theta_{\rm s}$. Let us assume that a UAV is located at $x$, and define $r_{\rm in}=h\tan({\theta_{\rm m}})$ and $r_{\rm out}=h\tan({\theta_{\rm s}})$. When $x\in b(o,r_{\rm in})$, where $b(o,r)$ denotes a two dimensional ball with radius $r$ centred at $o$, the UAV causes interference at the typical user with effective gain $G_{\rm m}$. When $x\in b(o,r_{\rm out})\setminus b(o,r_{\rm in})$, the co-channel UAV causes interference at the typical user with effective gain $G_{\rm s}$, and when $\|x\|>r_{\rm out}$, there is no interference from the UAV to the typical user.

We study the effect of correlation between time instants $0$ and $t$ on the joint success probability and the distribution of the number of interferers\footnote{Due to stationarity and ergodicity of our model, our results can be used for time instants $t_1$ and $t_2$, where $|t_2-t_1|=t$.}. Based on our system model, the total received interference at the typical user at time 0 is
\begin{IEEEeqnarray}{rCl}
	I_0 &=& \sum_{ x\in\Psi_0 }  h_x(0) \ell\left(\left(\|x\|^2+h^2\right)^{1/2}\right) \nonumber \\
	&& \times \left(G_{\rm m}\mathbf{1}(\|x\| \le r_{\rm in})+G_{\rm s}\mathbf{1}(r_{\rm in} < \|x\| \le r_{\rm out})\right), \nonumber
	\label{eq:I0}
\end{IEEEeqnarray}
where, according to Slivnyak's theorem, $\Psi_0=\Phi_0\setminus\{o\}$ is a PPP with intensity $\lambda$. $\mathbf{1}(.)$ denotes the indicator function. The total received interference at the typical user at time $t$ can be written as
\begin{IEEEeqnarray}{rCl}
	\IEEEeqnarraymulticol{3}{l}{I_t = \sum_{ x\in\Psi_0 } h_x(t) \Big[ M_x \ell\left(\left(\|x+\vec{\text v}_xt\|^2+h^2\right)^{1/2}\right)} \nonumber \\
	&&\times\> \left(G_{\rm m}\mathbf{1}(\|x+\vec{\text v}_xt\| \le r_{\rm in})+
	G_{\rm s}\mathbf{1}(r_{\rm in} < \|x+\vec{\text v}_xt\| \le r_{\rm out})\right) \nonumber \\
	&&+ \> (1-M_x) \ell\left(\left(\|x\|^2+h^2\right)^{1/2}\right) \nonumber \\
	&& \times \left(G_{\rm m}\mathbf{1}(\|x\| \le r_{\rm in})+G_{\rm s}\mathbf{1}(r_{\rm in} < \|x\| \le r_{\rm out})\right) \Big],
	\nonumber
	\label{eq:It} 
\end{IEEEeqnarray}
where $M_x$ is 1 when node $x$ is mobile, otherwise is zero; therefore, $M_x$ is a Bernoulli random variable with parameter $p$. When $M_x=1$, the projected location of the UAV onto $\mathbb{R}^2$ plane, at time $t$, is at distance $\|x+\vec{\text v}_xt\|=\left(\|x\|^2+v_x^2t^2-2\|x\|v_xt\cos(\theta_x)\right)^{1/2}$ from the origin.
Thus, the SINRs at the typical user at time instants 0 and $t$ are, respectively, as follows:
\begin{equation}
	\text{SINR}_0 = \frac{G_{\rm m}h_o(0)\ell(h)}{I_0+\sigma^2}, \quad
	\text{SINR}_t = \frac{G_{\rm m}h_o(t)\ell(h)}{I_t+\sigma^2}. \nonumber
\end{equation}

\section{Analytical Results}
\subsection{Joint Success Probability}
In the following theorem, we derive the joint success probability at time instants 0 and $t$.
\allowdisplaybreaks{
\begin{Theorem} \label{Thm2}
	Joint Success probability at time instants 0 and $t$ can be obtained by 
	\begin{IEEEeqnarray}{rCl}
		\IEEEeqnarraymulticol{3}{l}{\mathbb{P}\left\{ \textup{SINR}_0\ge T, \textup{SINR}_t \ge T \right\} =  \sum_{i,j=0}^{k-1} \frac{1}{i!j!} \times } \nonumber \\
		&&\> \left(\frac{\partial^i}{\partial s_1^i} \frac{\partial^j}{\partial s_2^j}
		e^{ \frac{Th^{\alpha}}{\Omega G_{\rm m}}(s_1+s_2)\sigma^2 } 
		\mathbb{E}\left[ e^{ \frac{Th^{\alpha}}{\Omega G_{\rm m}}(s_1 I_0+s_2 I_t) }  \right]  \right)_{\substack{s_1=-1\\s_2=-1}},	\nonumber 
		\label{Thm2:step1}
	\end{IEEEeqnarray}
	where
	\begin{IEEEeqnarray}{rCl}
		\IEEEeqnarraymulticol{3}{l}{
		\mathbb{E}\left[ e^{ \frac{Th^{\alpha}}{\Omega G_{\rm m}}(s_1 I_0+s_2 I_t) }  \right] = \exp\Bigg\{ -2\pi\lambda\int_0^{\infty}\Bigg[ 
		1 -} \nonumber 
	    \\
	 	&& \Bigg(1-s_1 \frac{T}{G_{\rm m}}\left(\frac{h^2}{h^2+x^2}\right)^{\alpha/2} \Big(G_{\rm m} \mathbf{1}(x\le r_{\rm in})+ \nonumber 
	 	\\
	 	&&G_{\rm s} \mathbf{1}(r_{\rm in}<x\le r_{\rm out})\Big) \Bigg)^{-k} \times \Bigg( \frac{p}{2\pi} \int_{0}^{\infty}\int_0^{2\pi}  \nonumber 
	 	\\
	 	&& \Bigg(1-s_2 \frac{T}{G_{\rm m}}\left(\frac{h^2}{h^2+x^2+v^2t^2-2xvt\cos\phi}\right)^{\alpha/2} \times \nonumber 
	 	\\
	 	&& \Big(G_{\rm m} \mathbf{1}(x^2+v^2t^2-2xvt\cos\phi\le r_{\rm in}^2)+ \nonumber 
	 	\\
	 	&&G_{\rm s} \mathbf{1}(r_{\rm in}^2 < x^2+v^2t^2-2xvt\cos\phi\le r_{\rm out}^2) \Big) \Bigg)^{-k} \nonumber
	 	f_V(v){\rm d}v{\rm d}\phi \nonumber
	 	\\
	 	&& + (1-p) \Bigg(1-s_2 \frac{T}{G_{\rm m}}\left(\frac{h^2}{h^2+x^2}\right)^{\alpha/2} \Big(G_{\rm m} \mathbf{1}(x\le r_{\rm in})+ \nonumber
	 	\\
	 	&&G_{\rm s} \mathbf{1}(r_{\rm in}<x\le r_{\rm out})\Big) \Bigg)^{-k} \Bigg) \Bigg] x {\rm d}x \Bigg\}. 
	 	\label{Thm2:step2}
	\end{IEEEeqnarray}
\end{Theorem}}
\allowdisplaybreaks{
\begin{IEEEproof}
	\begin{IEEEeqnarray}{rCl}
	\IEEEeqnarraymulticol{3}{l}{\mathbb{P}\left\{ \textup{SINR}_0\ge T, \textup{SINR}_t \ge T \right\} } \nonumber \\
	&=& \mathbb{P}\left\{ \frac{G_{\rm m}h_o(0)\ell(h)}{I_0+\sigma^2}\ge T, \frac{G_{\rm m}h_o(t)\ell(h)}{I_t+\sigma^2} \ge T \right\} \nonumber
	\\
	&\stackrel{(\text{a})}{=}& \mathbb{E}\Bigg[ \sum_{i,j=0}^{k-1} \frac{1}{i!j!} \left( \frac{Th^{\alpha}}{\Omega G_m} \right)^{i+j} \left(I_0+\sigma^2\right)^i \left(I_t+\sigma^2\right)^j \nonumber 
	\\
	&&  \times e^{-\frac{Th^{\alpha}}{\Omega G_m}  \left(I_0+\sigma^2\right)-\frac{Th^{\alpha}}{\Omega G_m}  \left(I_t+\sigma^2\right)} \nonumber
		\Bigg] \nonumber 
	\\
	&=& \mathbb{E}\left[ \sum_{i,j=0}^{k-1} \frac{1}{i!j!} \frac{\partial^i}{\partial s_1^i} \frac{\partial^j}{\partial s_2^j} 
	    e^{s_1\frac{Th^{\alpha}}{\Omega G_m}  \left(I_0+\sigma^2\right)+s_2\frac{Th^{\alpha}}{\Omega G_m}  \left(I_t+\sigma^2\right)} \right]_{\substack{s_1=-1\\s_2=-1}}, \nonumber 
	\\
	&\stackrel{(\text{b})}{=}& \sum_{i,j=0}^{k-1} \frac{1}{i!j!} \times \nonumber \\
	&& \left(\frac{\partial^i}{\partial s_1^i} \frac{\partial^j}{\partial s_2^j}
	e^{ \frac{Th^{\alpha}}{\Omega G_{\rm m}}(s_1+s_2)\sigma^2 } 
	\mathbb{E}\left[ e^{ \frac{Th^{\alpha}}{\Omega G_{\rm m}}(s_1 I_0+s_2 I_t) }  \right]  \right)_{\substack{s_1=-1\\s_2=-1}}, \nonumber 
	\end{IEEEeqnarray}
	where (a) is obtained by averaging over $h_o(0)$ and $h_o(t)$. The expectation in (b) can also be obtained by averaging with respect to $M_x$, $v_x$, $\theta_x$, fading power gains, and applying PGFL of PPP. 
\end{IEEEproof}}
Similar to \textbf{Theorem \ref{Thm2}}, we can derive the transmission success probability at time 0 (or $t$). 
\begin{IEEEeqnarray}{rCl}
	\IEEEeqnarraymulticol{3}{l}{\mathbb{P}\left\{ \textup{SINR}_0\ge T \right\}} \nonumber 
	\\ 
	&=&  \sum_{i=0}^{k-1} \frac{1}{i!} \left(\frac{\partial^i}{\partial s_1^i} 
	e^{ \frac{Th^{\alpha}}{\Omega G_{\rm m}}s_1\sigma^2 } 
	\mathbb{E}\left[ e^{ \frac{Th^{\alpha}}{\Omega G_{\rm m}}s_1 I_0 }  \right]  \right)_{s_1=-1}, \nonumber \\ \label{eq:success_at_0}
	\\
	\IEEEeqnarraymulticol{3}{l}{\mathbb{P}\left\{ \textup{SINR}_t\ge T \right\}} \nonumber 
	\\
	&=&  \sum_{j=0}^{k-1} \frac{1}{j!} \left(\frac{\partial^j}{\partial s_2^j}
	e^{ \frac{Th^{\alpha}}{\Omega G_{\rm m}}s_2\sigma^2 } 
	\mathbb{E}\left[ e^{ \frac{Th^{\alpha}}{\Omega G_{\rm m}}s_2 I_t }  \right]  \right)_{s_2=-1}, \nonumber \\ \label{eq:success_at_t}
\end{IEEEeqnarray}
where the expectations in \eqref{eq:success_at_0} and \eqref{eq:success_at_t} can be obtained by substituting $s_2=0$ and $s_1=0$ in \eqref{Thm2:step2}, respectively. To study the effect of correlation in mobile UAV networks, in the following, we derive the probability that the typical UAV has a successful transmission at time $t$, given that the transmission at 0 is unsuccessful, i.e.
\begin{IEEEeqnarray}{rCl}
	\IEEEeqnarraymulticol{3}{l}{
	\mathbb{P}\left\{ \textup{SINR}_t\ge T \mid \textup{SINR}_0< T \right\} = 
	\frac{ \mathbb{P}\left\{ \textup{SINR}_t\ge T , \textup{SINR}_0< T \right\} }{ \mathbb{P}\left\{ \textup{SINR}_0< T \right\} } } \nonumber \\
	&=& \frac{ \mathbb{P}\left\{ \textup{SINR}_t\ge T \right\} - \mathbb{P}\left\{ \textup{SINR}_t\ge T , \textup{SINR}_0\ge T \right\} }
		   { 1-\mathbb{P}\left\{ \textup{SINR}_0\ge T \right\} }. \nonumber 
\end{IEEEeqnarray}
Using Fortuin-Kasteleyn-Ginibre (FKG) inequality, one can show that $\mathbb{P}\left\{ \textup{SINR}_t\ge T \mid \textup{SINR}_0< T \right\} \le \mathbb{P}\left\{ \textup{SINR}_t\ge T\right\}$. 


\subsection{Distribution of the Number of Interferers}
First we calculate $\mathbb{P}\left\{\|x+\vec{\text v}_xt\| \le r\mid x,M_x=1 \right\}$, i.e.  the probability that the projected location of a mobile UAV at time $t$ is in $b(o,r)$ given its initial projected location is at  $x\in\mathbb{R}^2$. This probability, which is provided in the following lemma will be used to derive the distribution of the number of interferers.

\begin{Lemma} \label{Lemma1}
	The probability that the projected location of a mobile UAV at time $t$ is in $b(o,r)$ given its initial projected location is at  $x\in\mathbb{R}^2$, denoted by $F(r\mid \|x\|)$, can be derived as 
	\begin{IEEEeqnarray}{rCl}
		F(r\mid \|x\| ) &=& F_V\left(\frac{r-\|x\|}{t}\right) + \frac{1}{\pi} \int_{\frac{|\|x\|-r|}{t}}^{\frac{\|x\|+r}{t}} \nonumber \\
		&& \qquad \qquad  \arccos \left(\frac{\|x\|^2+v^2t^2-r^2}{2\|x\|vt}\right) f_V(v) {\rm d}v. \nonumber 
	\end{IEEEeqnarray}
\end{Lemma}
\allowdisplaybreaks{
	\begin{IEEEproof}
		\begin{IEEEeqnarray}{rCl}
			\IEEEeqnarraymulticol{3}{l}{\mathbb{P}\left\{ \left(\|x\|^2+v_x^2t^2-2\|x\|v_xt\cos(\theta_x)\right)^{1/2} \le r \mid x,M_x=1 \right\}} 
			\nonumber \\
			&=& \mathbb{P}\left\{ \cos(\theta_x)\ge \frac{\|x\|^2+v_x^2t^2-r^2}{2\|x\|v_xt} \mid x,M_x=1  \right\} \nonumber \\
			&=& \mathbb{E}_{v_x,\theta_x}\Bigg[ \mathbf{1} \left( \frac{\|x\|^2+v_x^2t^2-r^2}{2\|x\|v_xt} \le -1 \right) \nonumber \\
			&&+\> \mathbf{1} \left( -1 < \frac{\|x\|^2+v_x^2t^2-r^2}{2\|x\|v_xt} \le 1 \right) \nonumber \\
			&& \times \> \mathbf{1} \left(\cos(\theta_x)\ge \frac{\|x\|^2+v_x^2t^2-r^2}{2\|x\|v_xt}\right) \Bigg] \nonumber \\
			&=& \mathbb{E}_{v_x}\Bigg[  \mathbf{1} \left( v_x\le\frac{r-\|x\|}{t} \right) \nonumber \\
			&&+\> \mathbf{1} \left( \frac{|\|x\|-r|}{t} \le v_x\le\frac{\|x\|+r}{t} \right) \nonumber \\
			&& \times \frac{1}{\pi}\arccos \left(\frac{\|x\|^2+v_x^2t^2-r^2}{2\|x\|v_xt}\right) \Bigg] \nonumber 
		\end{IEEEeqnarray}
\end{IEEEproof}}
A UAV can cause interference at the typical user if its projected location is within $b(o,r_{\rm out})$. Therefore, at time 0, the number of interferers is a Poisson random variable with mean $\lambda\pi r_{\rm out}^2$. In the following, we derive the distribution of the number of interferers at time $t$ given that initially the number of interferers is $m$. 

\begin{Theorem} \label{Thm1}
	The probability that there are $n$ interferers at time $t$ given that initially the number of interferers is $m$ is 
	\begin{IEEEeqnarray}{rCl}
		\IEEEeqnarraymulticol{3}{l}{\mathbb{P}\left\{\Psi_t \left(b(o,r_{\rm out})\right)=n \mid \Psi_0 \left(b(o,r_{\rm out})\right)=m \right\}}
		\nonumber \\
		&=& \frac{1}{n!}\sum_{i=0}^{\min(n,m)} \binom{n}{i} \frac{m!}{(m-i)!} p^{m-i} \nonumber \\
		&& \times\> \left( 1- \int_0^{r_{\rm out}} F(r_{\rm out}\mid x) \frac{2x}{r_{\rm out}^2} {\rm d}x \right)^{m-i} 
		\nonumber \\ 
		&&\times\> \left( p\int_0^{r_{\rm out}} F(r_{\rm out}\mid x) \frac{2x}{r_{\rm out}^2} {\rm d}x +1-p\right)^{i} \nonumber \\
		&&\times\> \left(2\pi \lambda p\int_{r_{\rm out}}^{\infty} F(r_{\rm out}\mid x) x {\rm d}x \right)^{n-i} \nonumber \\
		&&\times \> \exp\left\{-2\pi \lambda p\int_{r_{\rm out}}^{\infty} F(r_{\rm out}\mid x) x {\rm d}x \right\}, \nonumber
	\end{IEEEeqnarray}
	where $\Psi_t=\Phi_t\setminus\{o\}$.  
\end{Theorem}
\allowdisplaybreaks{
\begin{IEEEproof}
	\textbf{Theorem \ref{Thm1}} can be obtained by using
	\begin{multline}
	\mathbb{P}\left\{\Psi_t \left(b(o,r_{\rm out})\right)=n \mid \Psi_0\left(b(o,r_{\rm out})\right)=m \right\} =\\ \frac{1}{n!} \frac{{\rm d}^n}{{\rm d}s^n}\mathbb{E}\left[ s^{\Psi_t\left(b(o,r_{\rm out})\right)} \mid \Psi_0\left(b(o,r_{\rm out})\right)=m \right] \mid _ {s=0}.
	\label{thm1:step1}
	\end{multline}
	Given $\Psi_0\left(b(o,r_{\rm out})\right)=m$, we can consider $\Psi_0$ as a superposition of two independent point processes: i) a uniform BPP with $m$ number of points in $b(o,r_{\rm out})$, and ii) a PPP with intensity function $\lambda\mathbf{1}(\|x\|>r_{\rm out})$. Therefore, we have
	\begin{IEEEeqnarray}{rCl}
		\IEEEeqnarraymulticol{3}{l}{\mathbb{E}\left[ s^{\Psi_t\left(b(o,r_{\rm out})\right)} \mid \Psi_0\left(b(o,r_{\rm out})\right)=m \right]} \nonumber \\
		&=&  \mathbb{E}\Bigg[ \prod_{x\in \Psi_{0}\cap b(o,r_{\rm out})} 
		\Big( M_x s^{\mathbf{1}\left(\|x+\vec{\text v}_xt\|\le r_{\rm out}\right)} \nonumber \\
		&& \qquad \quad +\> (1-M_x) s^{\mathbf{1}\left(\|x\|\le r_{\rm out}\right)}  \Big) \mid \Psi_0\left(b(o,r_{\rm out})\right)=m  \Bigg] \nonumber \\
		&&\times \> \mathbb{E}\Bigg[ \prod_{x\in \Psi_{0}\cap b^c(o,r_{\rm out})} 
		\Big( M_x s^{\mathbf{1}\left(\|x+\vec{\text v}_xt\|\le r_{\rm out}\right)} \nonumber \\
		&&\qquad \qquad \qquad \qquad \qquad \qquad +\> (1-M_x) s^{\mathbf{1}\left(\|x\|\le r_{\rm out}\right)}  \Big)  \Bigg] \nonumber \\
		&\stackrel{\text{(a)}}{=}& \left( p\left(1+(s-1)\int_0^{r_{\rm out}} F(r_{\rm out} \mid x )\frac{2x}{r^2_{\rm out}} {\rm d}x \right) + (1-p)s \right)^m \nonumber \\
		&& \times \exp\left\{ -2\pi\lambda p (1-s) \int_{r_{\rm out}}^{\infty} F(r_{\rm out} \mid x) x {\rm d}x \right\},
		\label{thm1:step2}
	\end{IEEEeqnarray}
	where (a) is obtained by using probability generating functional (PGFL) of PPP. Finally, \textbf{Theorem \ref{Thm1}} is obtained by substituting \eqref{thm1:step2} in \eqref{thm1:step1}, and employing the general Leibniz rule for the $n$-th derivative of a product of two functions.
\end{IEEEproof}}
We can also show that mean number of arrival of interferers (mean number of UAVs that only cause interference at the typical user at time $t$) is equal to 
\begin{multline}
\mathbb{E}\left[ \sum_{x\in\Psi_{0}\cap b^c(o,r_{\rm out})} M_x\mathbf{1}\left(\|x+\vec{\text v}_xt\|\le r_{\rm out}\right) \right] 
\\= 2 \pi \lambda p \int_{r_{\rm out}}^{\infty} F(r_{\rm out} \mid x) x {\rm d}x, \nonumber
\end{multline}
and mean number of departure of interferers (mean number of UAVs that only cause interference at the typical user at time $0$) is
\begin{IEEEeqnarray}{rCl} 
\IEEEeqnarraymulticol{3}{l}{ 
\mathbb{E}\left[ \sum_{x\in\Psi_{0}\cap b(o,r_{\rm out})} M_x\mathbf{1}\left(\|x+\vec{\text v}_xt\|> r_{\rm out}\right) \mid \Psi_0\left(b(o,r_{\rm out})\right)=m \right] 
} \nonumber
\\
&=&  mp \left(  1- \int_0^{r_{\rm out}} F(r_{\rm out} \mid x) \frac{2x}{r_{\rm out}^2} {\rm d}x \right). \nonumber
\end{IEEEeqnarray}
As derived in \textbf{Theorem \ref{Thm1}}, due to mobility of UAVs, the number of interferers is different in different time slots, and depending on the network parameters the mean number of interferers may increase or decrease. 
When the mean number of interferers increases, i.e. the mean number of arrivals is greater than mean number of departures, retransmitting an unsuccessfully received packet may not be helpful, while when the mean number of interferers decreases, if there is no stringent latency constraint, the typical UAV should postpone the transmission to   increase reliability. 

\section{Numerical and Simulation Results}

\begin{figure}[ht]
	\parbox[b]{.5\textwidth}{%
		\centerline{\subfigure[$t=1$.]
			{\epsfig{file=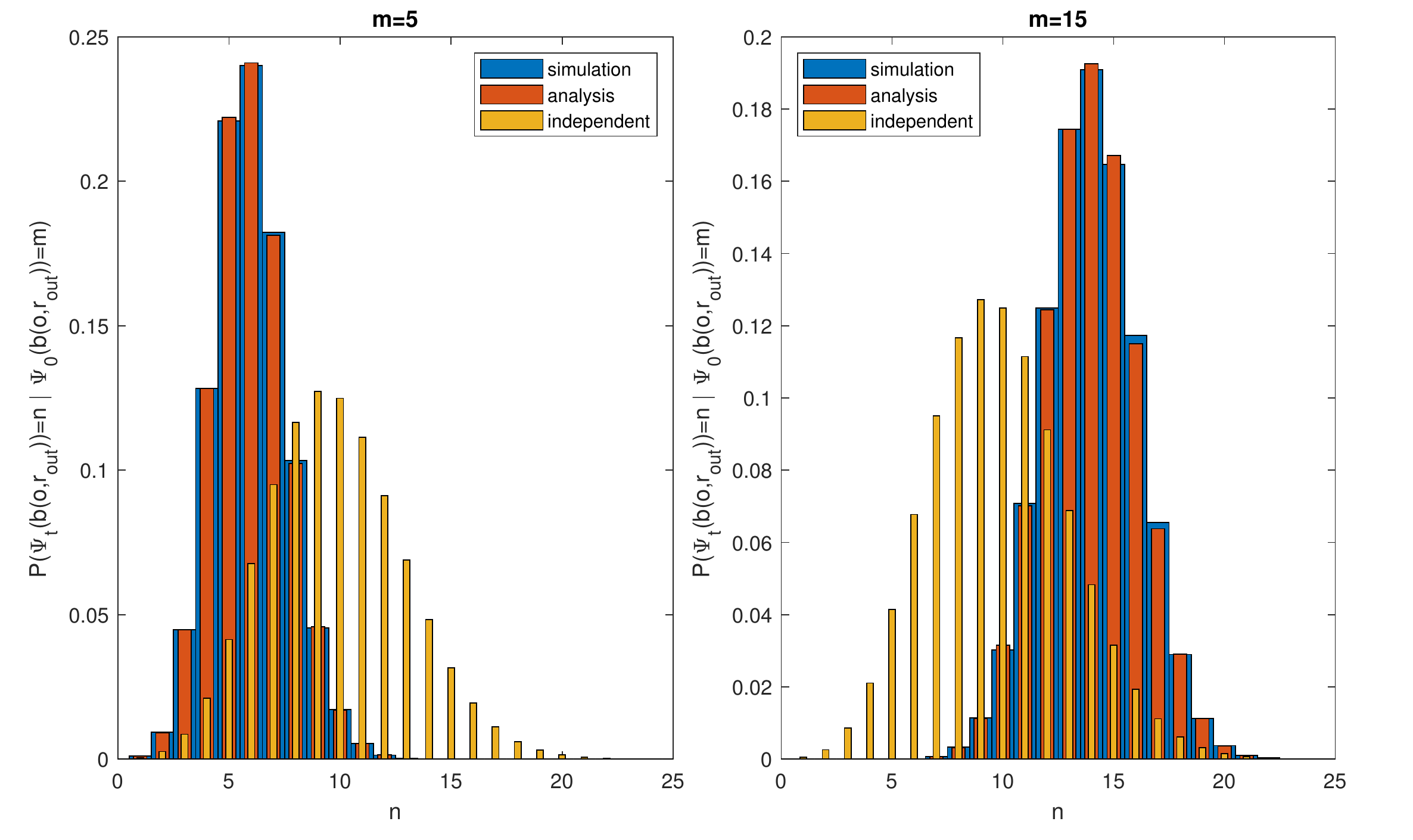, width=.4\textwidth}}}} 
	\parbox[b]{.5\textwidth}{%
		\centerline{\subfigure[$t=5$.]
			{\epsfig{file=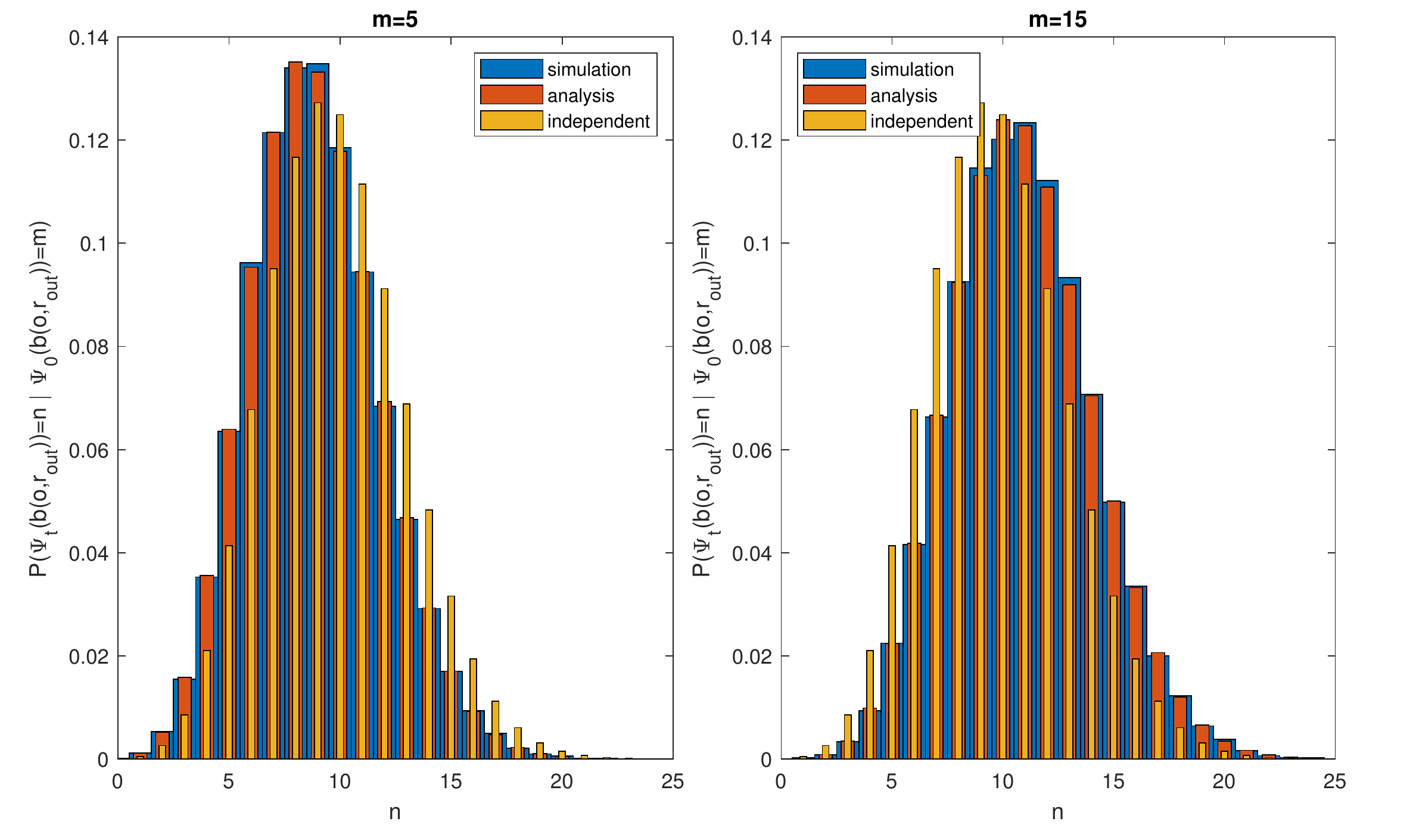, width=.4\textwidth}}}}
	\caption{Distribution of the number of interferers at time $t$, given, at time 0, the number of interferers is $m$ (for $\lambda=0.005$, $r_{\rm in}=15$, $r_{\rm out}=25$, $p=0.8$). For all UAVs, $v=10$ and their movement directions are uniformly distributed in $[0,2\pi]$.}
	\label{fig:N_I}
\end{figure}

Distribution of the number of interferers at $t=1$ and $t=5$, given, at time 0, the number of interferers is $m$ is illustrated in \figref{fig:N_I}(a) and (b), respectively, for $m=5$ and $m=15$. When $m=5$, at $t=1$ ($t=5$), mean number of arrivals is about 2 (8) and mean number of departures is about 1 (4). Thus, retransmitting an unsuccessfully received packet may not be helpful in this scenario. On the other hand, for $m=15$, at $t=1$ ($t=5$), the mean number of arrivals and departures are about 2 (8) and 3 (12), respectively. Therefore, when there is no strict latency constraint,  postponing the transmission for time $t$ provides a higher reliability. Moreover, to study the effect of correlation on the distribution of the number of interferers, in \figref{fig:N_I}, we compare the results with the independent scenario, where the distribution of the number of interferers follows a Poisson random variable with mean $\lambda \pi r_{\rm out}^2$. As can be seen, the correlation decreases as the time gap $t$ increases.

\begin{figure}[ht]
	\parbox[b]{.5\textwidth}{%
		\centerline{\subfigure[$m=5$.]
			{\epsfig{file=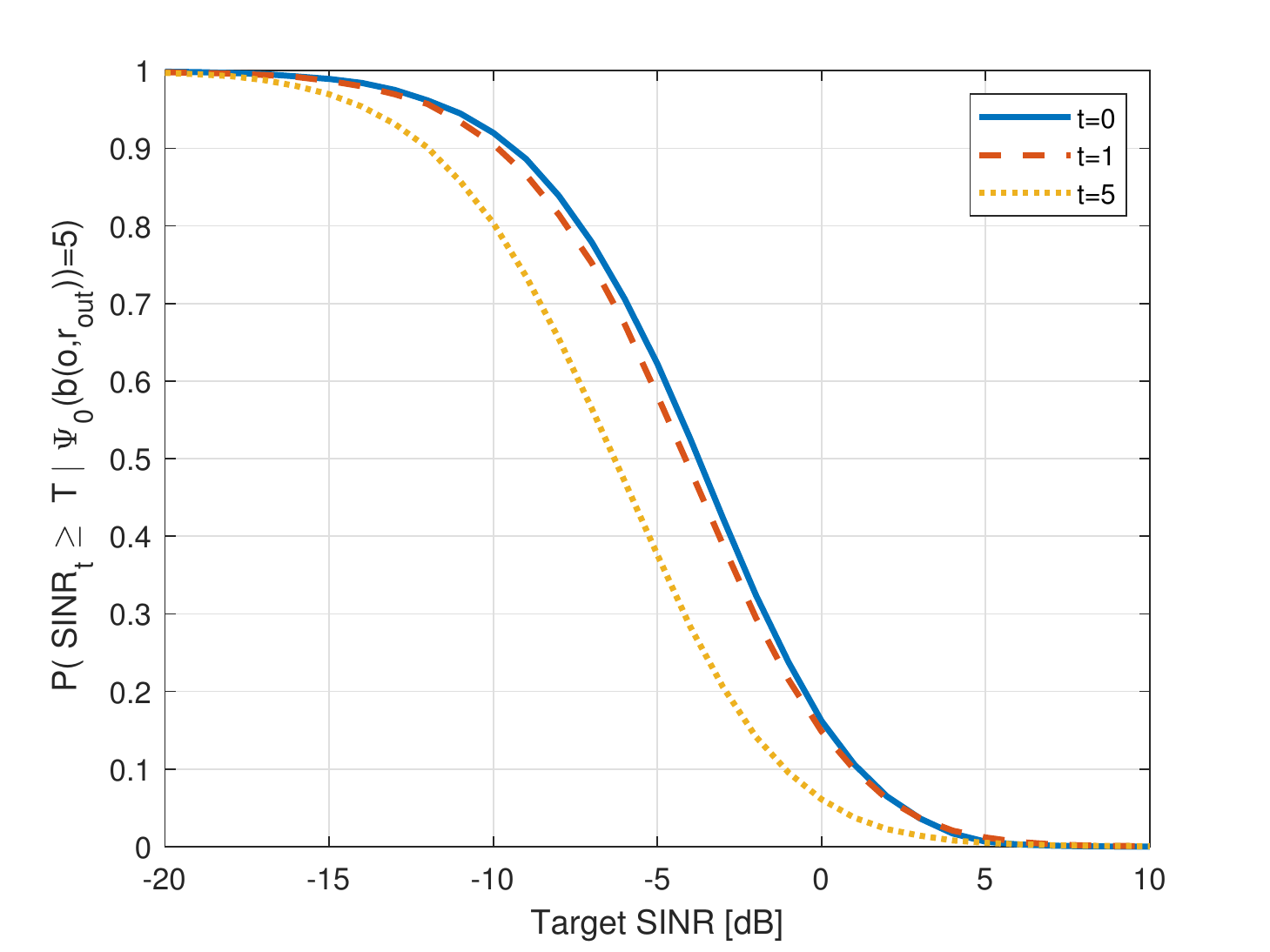, width=.4\textwidth}}}} 
	\parbox[b]{.5\textwidth}{%
		\centerline{\subfigure[$m=15$.]
			{\epsfig{file=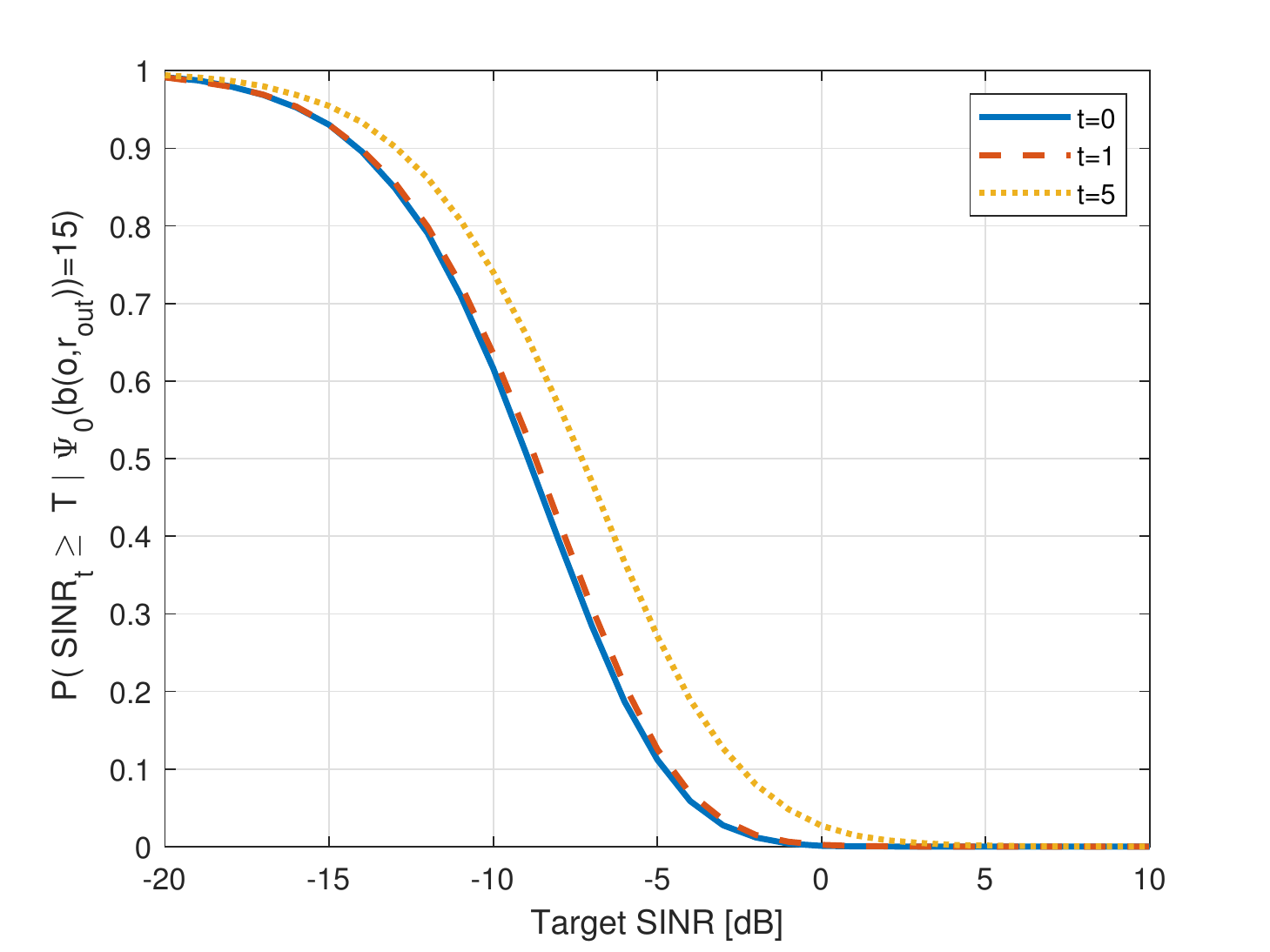, width=.4\textwidth}}}}
	\caption{Conditional success probability at time $t$, given, at time 0, the number of interferers is $m$ (for $\lambda=0.005$, $r_{\rm in}=15$, $r_{\rm out}=25$, $h=50$, $G_{\rm m}=2$, $G_{\rm s}=0.5$, $k=2$, $\Omega=1/2$, $p=0.8$, $\sigma^2=10^{-10}$, and $\alpha=4$). For all UAVs, $v=10$ and their movement directions are uniformly distributed in $[0,2\pi]$.}
	\label{fig:ConditionalCoverage}
\end{figure}

\figref{fig:ConditionalCoverage} illustrates the variation in conditional success probability at time $t$ with different target SINR, given that, at time 0, the number of interferers is $m$. These results are obtained from simulations. For $m=5$, the conditional success probability decreases as $t$ increases, while for $m=15$ the conditional success probability increases as $t$ increases.

\begin{figure}
	\centering
	\includegraphics[width=.4\textwidth]{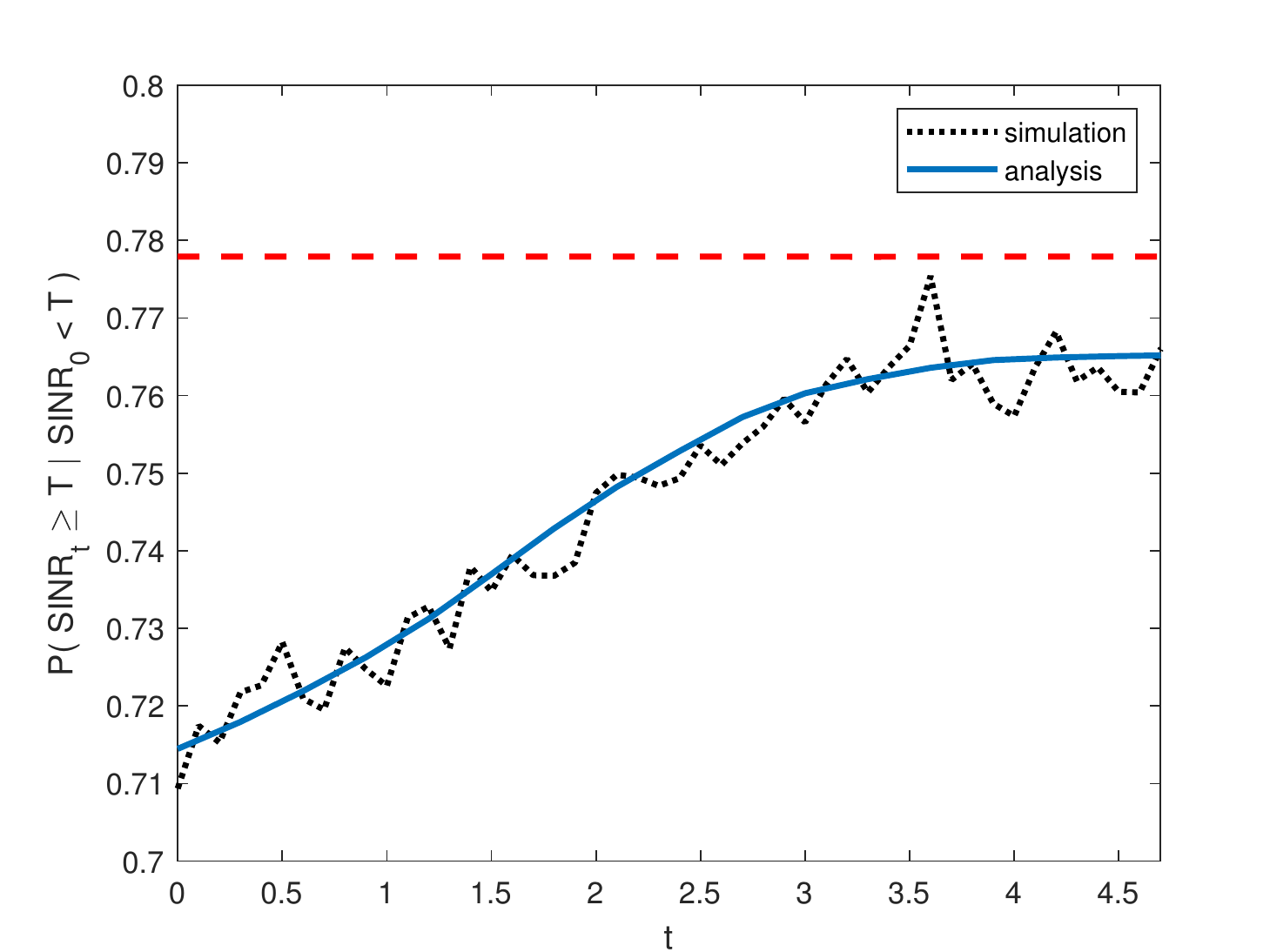}
	\caption{Effect of temporal correlation in mobile UAV networks (for $\lambda=0.005$, $r_{\rm in}=15$, $r_{\rm out}=25$, $h=50$, $G_{\rm m}=2$, $G_{\rm s}=0.5$, $k=2$, $\Omega=1/2$, $p=0.8$, $\sigma^2=10^{-10}$, $\alpha=4$, and $T=-10{\rm dB}$). For all UAVs, $v=10$ and their movement directions are uniformly distributed in $[0,2\pi]$. }
	\label{fig:retransmission}	
\end{figure}

We also study the effect of temporal correlation in mobile UAV networks in \figref{fig:retransmission}. Specifically, we have illustrated the success probability at time $t$ given initial transmission attempt at time $0$ fails. As the time gap between the two transmission attempts, denoted by $t$, increases, success probability increases since the correlation between the two time instants decreases. We also compare our results with the independent scenario (shown by the red dashed line). As is evident, as $t$ increases we can ignore the correlation.


\section{Conclusion}
We have studied the effect of temporal correlation on the distribution of the number of interferers and success probability in mobile UAV networks. 
We have shown that in some scenarios, this correlation can be exploited by postponing a transmission in order to  increase the success probability (or reliability). 
As the time gap between two transmission attempts increases the effect of correlation decreases. 
Our results can be used to optimize the error recovery protocols (e.g. ARQ protocols) and study the rate-reliability-latency tradeoffs in UAV networks. In this letter, we have only considered the effect of temporal correlation on the downlink performance of a user served by a static UAV in a network that includes both static and mobile UAVs. For future work, we can similarly study the effect of temporal correlation on the performance of a user served by mobile UAVs in the same network.

\IEEEpeerreviewmaketitle
\bibliographystyle{IEEEtran}
\bibliography{IEEEabrv,Bibliography}

\end{document}